\documentclass[manuscript,screen,nonacm]{acmart}

\setcopyright{none}
\renewcommand\footnotetextcopyrightpermission[1]{}
\makeatletter
\def\@ACM@checkaffil{}
\@twosidefalse
\@mparswitchfalse
\makeatother
\usepackage[T1]{fontenc}
\usepackage{amsmath}
\usepackage{booktabs}
\usepackage{subcaption}
\usepackage{graphicx}
\usepackage{float}
\usepackage{algorithm}
\usepackage{algpseudocode}

\newcommand{\cellocuttableformat}{%
  \small
  \setlength{\tabcolsep}{7pt}%
  \renewcommand{\arraystretch}{1.08}%
}

\title{CelloCut: Constructive Watertight Remeshing via Tetrahedral Cell Cuts}

\author{Xuan Yang}
\authornote{Equal contributions.}
\email{xuanyang2005@gmail.com}
\affiliation{%
  \institution{Nanjing University}
  \city{Suzhou}
  \state{Jiangsu}
  \country{China}}

\author{Yuhang Zeng}
\authornotemark[1]
\email{yuhaaa104@gmail.com}
\affiliation{%
  \institution{Nanjing University}
  \city{Suzhou}
  \state{Jiangsu}
  \country{China}}

\author{Dinglong Fang}
\authornotemark[1]
\email{vincent8510182346@outlook.com}
\affiliation{%
  \institution{Nanjing University}
  \city{Suzhou}
  \state{Jiangsu}
  \country{China}}

\author{Guochuan Tang}
\email{t0xuanshou@gmail.com}
\affiliation{%
  \institution{Nanjing University}
  \city{Suzhou}
  \state{Jiangsu}
  \country{China}}

\author{Jiaju Jiang}
\email{jiangjjsgs@js.chinamobile.com}
\affiliation{%
  \institution{China Mobile Zijin (Jiangsu) Innovation Research Institute Co., Ltd.}
  \city{Nanjing}
  \state{Jiangsu}
  \country{China}}

\author{Ben Li}
\email{liben@js.chinamobile.com}
\affiliation{%
  \institution{China Mobile Zijin (Jiangsu) Innovation Research Institute Co., Ltd.}
  \city{Nanjing}
  \state{Jiangsu}
  \country{China}}

\author{Wei Zhou}
\email{zhouwei9@js.chinamobile.com}
\affiliation{%
  \institution{China Mobile Zijin (Jiangsu) Innovation Research Institute Co., Ltd.}
  \city{Nanjing}
  \state{Jiangsu}
  \country{China}}

\author{Xiao-Xiao Long}
\authornote{Corresponding author.}
\email{xxlong@nju.edu.cn}
\affiliation{%
  \institution{Nanjing University}
  \city{Suzhou}
  \state{Jiangsu}
  \country{China}}

\author{Cheng Lin}
\authornotemark[2]
\email{chenglin@must.edu.mo}
\affiliation{%
  \institution{Macau University of Science and Technology}
  \city{Macau}
  \country{China}}

\begin{teaserfigure}
  \centering
  \includegraphics[width=0.9\linewidth]{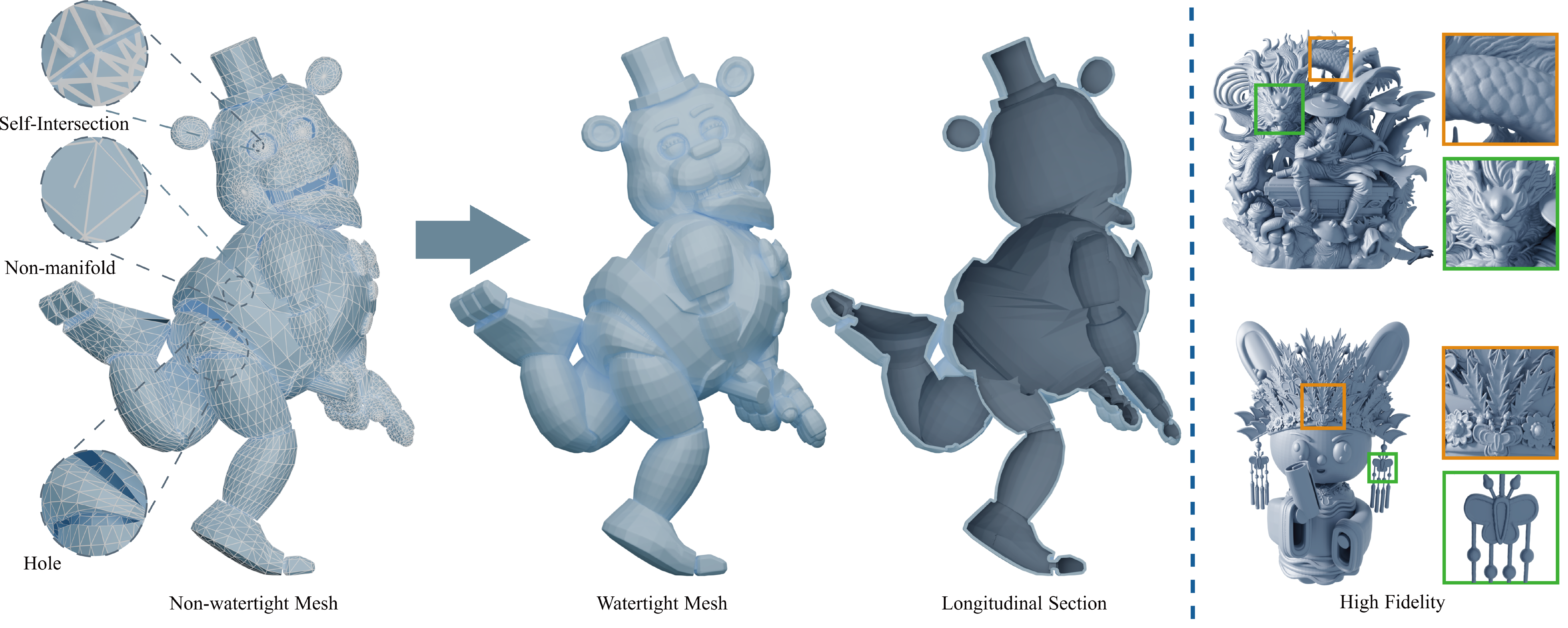}
  \caption{Starting from a raw mesh riddled with structural holes, self-intersections, and non-manifold edges (left), CelloCut converts the input into a compact and strictly watertight solid (middle). The resulting volumetric partition induces a globally consistent interior--exterior interpretation, as shown in the longitudinal section, and supports reconstruction of fine-scale boundary detail in the final surface (right).}
  \label{fig:teaser}
\end{teaserfigure}

\begin{abstract}
Watertight remeshing aims to recover a surface that induces a globally consistent interior--exterior partition of 3D space.
However, for meshes with complex topology, single-layer structures, or large missing regions, inferring such a partition from local surface geometry is inherently ambiguous.
As a result, existing methods often produce surface-accurate yet volumetrically inconsistent reconstructions, $e.g$., closely spaced double shells.
The key insight of this work is that watertight remeshing should be treated as a volumetric partitioning problem rather than a surface-level repair task. To this end, we propose CelloCut, a constructive framework that formulates watertight conversion as a binary labeling problem over a Delaunay tetrahedral partition of space. We solve this via graph-cut energy minimization with one-sided constraints that preserve proxy-supported interior evidence and weighted interface penalties that discourage unsupported newly introduced boundaries. By computing a globally consistent volumetric partition, CelloCut guarantees a strictly watertight output by construction and strongly suppresses pseudo-watertight artifacts such as double shells, even under severe topological defects.
Experimental results on two newly introduced challenging benchmarks, CelloScan and CelloFill, as well as standard ModelNet10 dataset, demonstrate that CelloCut significantly outperforms state-of-the-art methods, particularly in handling complex topologies and single-layer structures, producing compact and volumetrically consistent solid reconstructions.
The project page is available at \url{https://rangeryx-66.github.io/CelloCut/}.
\end{abstract}

\ccsdesc[500]{Computing methodologies~Mesh geometry models}
\ccsdesc[300]{Computing methodologies~Mesh models}

\keywords{watertight remeshing, volumetric partitioning, tetrahedral mesh, graph cut, mesh repair}

\begin{document}
\maketitle

\section{Introduction}
Watertight surface reconstruction is fundamental to 3D geometry processing and underpins applications such as physical simulation\cite{DBLP:conf/cvpr/XieZQLF0J24,DBLP:conf/iros/TodorovET12,DBLP:conf/nips/MakoviychukWGLS21,DBLP:conf/cvpr/XiangQMXZLLJYWY20,DBLP:conf/iclr/HuALSCRD20,DBLP:conf/iclr/GuXLLLMTTWYYXHC23,DBLP:conf/nips/FreemanFRGMB21,DBLP:conf/nips/GanSAMSTFKBHSKW21}, computational fabrication\cite{DBLP:journals/cgf/LivesuEMLA17,DBLP:journals/tog/ZhangFHDLKW22,DBLP:journals/tog/MartinezDL16,DBLP:journals/tog/ZhaoGHGCTBZCC16,DBLP:journals/tog/JacobsonKS13,DBLP:journals/tog/JourdanHSML23,DBLP:journals/tog/ZhongZLYLCL23}, and 3D generative model training\cite{DBLP:conf/cvpr/XiangLXDWZC0Y25,hunyuan3d2025hunyuan3d,DBLP:conf/nips/WuLZZ00C024,wu2025direct3d,DBLP:journals/tog/ZhangTNW23,lai2025lattice,seed2025seed3d,hunyuan3d22025tencent,yang2024hunyuan3d,jia2025ultrashape,li2025triposg}. True watertightness requires a globally consistent partition of space into interior and exterior regions, yet this requirement is often ignored by surface-level reconstruction objectives and evaluation metrics.

In practice, many real-world meshes from scanning, manual modeling, or generative pipelines do not uniquely define an interior volume. Complex topology, near-zero-thickness structures, and large missing regions introduce intrinsic ambiguity, making occupancy inference from surface signals fundamentally ill-posed. Prior state-of-the-art methods exhibit distinct limitations in addressing these challenges.  Projection-based ManifoldPlus \cite{huang2020manifoldplus} ensures watertightness but produces visual spikes due to insufficient regularization. VolumeMesher \cite{DBLP:journals/tog/DiazziA21} enforces closed outputs, but lacks vertex-manifoldness guarantees, leading to collapsed thin geometry and singularities. Implicit pipelines like Dora \cite{DBLP:conf/cvpr/ChenZLLLLLLFT25} rely on UDF dilation, which fails to bridge large gaps and causes double shells, while Craftsman \cite{DBLP:conf/cvpr/LiLYCLCTL25} suffers from aliasing and limited robustness. As a result, these methods may achieve low surface error under standard metrics while producing volumetrically inconsistent results—such as double shells or leaky solids—which break downstream volumetric and physical pipelines. In such cases, the goal is often not exact surface restoration, but robust conversion to a compact, single-shell, and volumetrically consistent solid with a globally consistent interior–exterior interpretation.

We therefore view watertight conversion of defective meshes not as repairing a surface, but as selecting a conservative solid interpretation from incomplete and topologically ambiguous boundary evidence. This distinction is crucial: once holes, self-intersections, and near-zero-thickness structures are present, the input surface no longer defines a unique interior volume, and surface-level fidelity alone becomes an insufficient objective. What is needed instead is a formulation that reasons directly about interior and exterior regions and resolves ambiguity at the volumetric level. Based on this view, we propose CelloCut, a constructive framework that embeds the defective input into a tetrahedral discretization of space and solves for a globally consistent interior-exterior partition over volumetric cells. Our formulation is guided by a conservative thickened proxy that provides reliable interior support, one-sided constraints that preserve that support while leaving ambiguous space free to be optimized, and fill-aware interface regularization that favors compact completions with minimal unsupported boundary. The final watertight surface is then extracted as the interface induced by the optimized volumetric partition.

Although our optimization also uses graph cuts over a tetrahedral partition, our problem setting is fundamentally different from classical graph-cut volumetric reconstruction. Existing graph-cut methods typically recover surfaces from observations such as point samples or visibility cues, whereas our input is a defective triangle mesh whose inside-outside relation is itself ambiguous. Accordingly, our objective is not to fit observations with a surface, but to compute a conservative volumetric interpretation that yields a strictly watertight solid by construction. A more detailed discussion of this distinction is provided in Sec.~\ref{sec:relations}.

Experiments on two new challenging benchmarks, CelloScan and CelloFill, as well as the standard ModelNet10 dataset \cite{DBLP:conf/cvpr/WuSKYZTX15}, demonstrate robust watertight reconstruction under severe topological ambiguity, single-layer structures, and large missing regions, significantly outperforming state-of-the-art methods.


Our formulation is driven by three observations about watertight processing under topological ambiguity, each of which directly shapes the design of the method:
\begin{itemize}
    \item When a defective mesh no longer defines a unique inside-outside relation, watertight conversion should be posed as selecting a solid interpretation rather than repairing a surface. In such cases, the goal is not exact surface restoration, but a globally consistent volumetric hypothesis.
    \item The available geometric evidence is fundamentally asymmetric. Interior regions supported by a conservative thickened proxy are reliable enough to preserve, whereas the remaining space is ambiguous and should remain free to be optimized. This asymmetry is what makes conservative yet effective hole filling possible.
    \item Once watertight conversion is cast as asymmetric volumetric partitioning, unsupported completion hypotheses should be more expensive than proxy-supported interfaces. This biases the solution toward minimal unsupported boundary, favoring compact single-shell completion over fragmented or double-shell-like alternatives.
\end{itemize}

\begin{figure*}[t]
  \centering

  \includegraphics[width=\textwidth]{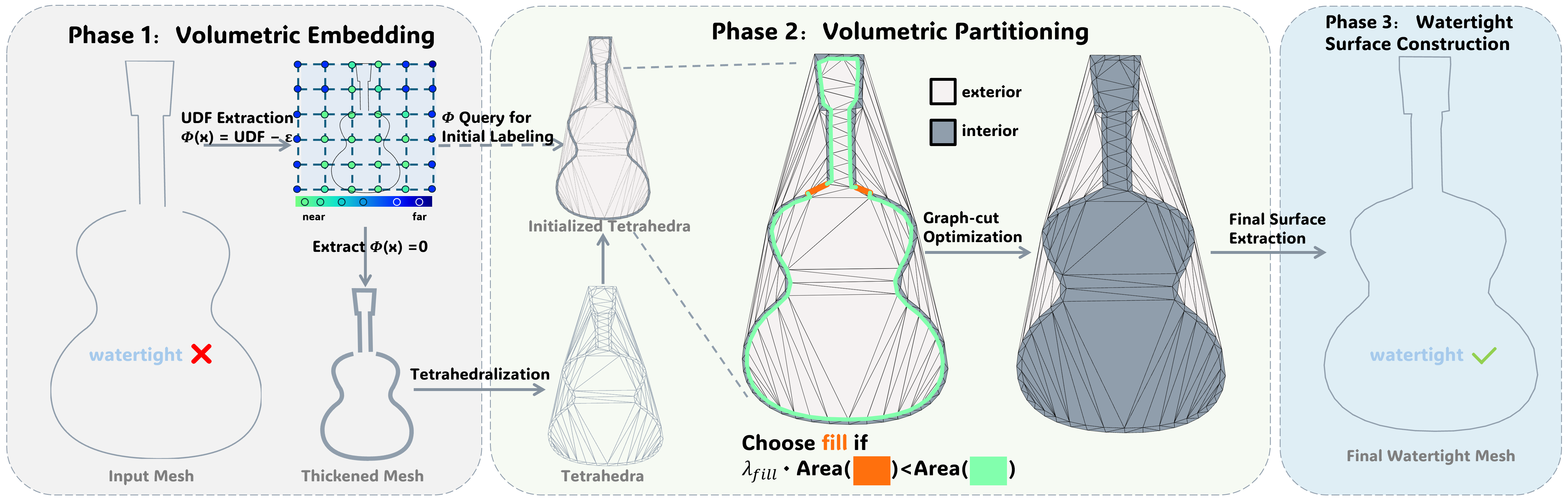}

  \caption{Overview of CelloCut. CelloCut treats watertight remeshing as a volumetric partitioning problem rather than a surface repair task, and proceeds in three stages: 1) embedding the input non-watertight mesh into a volumetric representation via UDF-based surface thickening and tetrahedralization; 2) resolving topological ambiguity by graph-cut optimization of a constrained volumetric labeling with area-based regularization; and 3) extracting a watertight surface by construction from the optimized volumetric partition. This figure is shown in 2D for clarity and illustrates the underlying principle rather than geometric accuracy. }

  \label{fig:overview_pipeline} 
\end{figure*}

\section{Related Work}

Transforming imperfect real-world geometry into watertight, simulation-ready assets involves distinct challenges in robustness and fidelity. Existing methods generally fall into two categories: surface-based repair and volumetric reconstruction.



\subsection{Surface-based Repair Strategies}
Surface-based repair methods directly modify the input mesh to restore manifoldness and watertightness, typically by detecting non-manifold elements and applying local topology-driven operations such as mesh zippering and boundary-based hole filling \cite{DBLP:journals/vc/Attene10,DBLP:journals/tvcg/GueziecTLH01,attene2018exact,DBLP:conf/siggraph/TurkL94,DBLP:conf/sgp/Liepa03,DBLP:journals/cagd/BarequetS95}.
Recent visual-measure-based approaches further leverage ray-tracing cues to guide surface closing under the assumption of visually correct geometry \cite{DBLP:journals/tvcg/ZhengGPLWWW24}.
Closely related to this line of work, recent robust Boolean, mesh-arrangement, and exact solid-modeling methods focus on resolving intersections and extracting topologically valid solids through exact predicates, exact constructions, or local arrangement computation, e.g., Interactive and Robust Mesh Booleans \cite{DBLP:journals/tog/CherchiPAL22}, EMBER \cite{DBLP:journals/tog/TrettnerNK22}, Exact and Efficient Intersection Resolution for Mesh Arrangements \cite{DBLP:journals/tog/GuoF24}, and Exact Predicates, Exact Constructions and Combinatorics for Mesh CSG \cite{DBLP:journals/tog/Levy25}.
While these methods are highly effective when valid solids can be recovered through exact intersection resolution and combinatorial arrangement construction, they are less suitable for ambiguous single-sided or large-gap cases where missing geometry must be inferred rather than exactly resolved.
More generally, surface-based methods rely primarily on local surface configurations or exact surface combinatorics, and lack an explicit global volumetric formulation of interior--exterior consistency, limiting their robustness to severe topological ambiguity and incomplete geometry.

\subsection{Volumetric Reconstruction and Remeshing}
To achieve robustness, strictly watertight methods typically reformulate the problem as extracting an interface from an intermediate volumetric representation.

\noindent\textbf{Grid-based Occupancy and Solidification.}
Grid-based methods for watertight remeshing mainly differ in how interior--exterior occupancy is inferred from volumetric discretizations.
Manifold and ManifoldPlus \cite{huang2018robust,huang2020manifoldplus} infer occupancy via voxelization and flood-fill, while Dora \cite{DBLP:conf/cvpr/ChenZLLLLLLFT25} relies on UDF-based surface thickening.
Related UDF-based remeshing methods also generate watertight outputs by extracting offset surfaces from distance fields and subsequently optimizing the extracted mesh, as in Robust Low-Poly Meshing \cite{DBLP:journals/tog/ChenPWVG23}; similar mesh-to-volume-to-mesh pipelines are also explored in recent low-poly optimization systems such as PaMO \cite{DBLP:journals/cgf/OhYWSXLS25}.
CLAY \cite{DBLP:journals/tog/ZhangWZQPJYXY24} and CraftsMan3D \cite{DBLP:conf/cvpr/LiLYCLCTL25,DBLP:journals/tog/JacobsonKS13} further incorporate visibility and winding-number tests.
Sparc3D \cite{li2025sparc3d} combines UDF-based surface extraction with flood-fill labeling; although these approaches guarantee watertightness, they often struggle to repair large openings or missing regions and may introduce spurious fine-scale artifacts, such as thin spikes and blocky voxel-like structures.

\noindent\textbf{Optimization on Adaptive Spatial Discretizations.}
A line of work enforces global interior--exterior consistency by formulating watertight reconstruction as volumetric partitioning on adaptive discretizations.
Early methods leverage Delaunay tessellations to extract watertight surfaces via natural neighbor interpolation \cite{DBLP:journals/comgeo/BoissonnatC02} or graph-cut-based inside/outside labeling on adaptive tetrahedral complexes \cite{DBLP:conf/iccv/LabatutPK07,10.5555/1281957.1281963,DBLP:journals/cgf/LabatutPK09}. In particular, Hornung and Kobbelt \cite{10.5555/1281957.1281963} reconstruct watertight surfaces from unoriented point clouds by extracting a minimum-cut surface from an unsigned-distance confidence field on a volumetric grid, while Labatut et al. \cite{DBLP:conf/iccv/LabatutPK07,DBLP:journals/cgf/LabatutPK09} formulate reconstruction as an energy minimization over Delaunay tetrahedra, where graph cuts recover a globally optimal inside/outside labeling under visibility- and data-driven terms.
Subsequent approaches build on Delaunay refinement and variational optimization theory \cite{DBLP:journals/comgeo/Shewchuk02,DBLP:journals/tog/AlliezCYD05} to enable robust adaptive tetrahedralizations (e.g., TetWild and fTetWild \cite{DBLP:journals/tog/HuSWZP20,Hu:2018:TMW:3197517.3201353}), while related work emphasizes numerically robust solid modeling via iterative space partitioning and cell labeling \cite{DBLP:journals/tog/DiazziA21}. Our approach follows this general paradigm but differs in both problem setting and optimization design. Existing graph-cut-based volumetric reconstruction methods mainly recover surfaces from point samples, range scans, or multi-view observations using visibility- or data-driven terms, whereas we address watertight conversion of defective triangle meshes whose inside--outside relation is fundamentally ambiguous. Accordingly, our method does not seek to recover a surface directly from observations; instead, it constructs a conservative thickened proxy and solves a constrained volumetric labeling problem with one-sided interior anchoring and fill-aware interface regularization, yielding a watertight solid completion rather than a generic surface reconstruction.

\subsection{Relation to Graph-Cut Volumetric Reconstruction}\label{sec:relations}
Our method uses the same optimization primitive as classical graph-cut volumetric reconstruction, but it addresses a fundamentally different problem setting. Existing graph-cut reconstruction methods typically start from observations such as point samples, visibility cues, or range scans, and estimate inside-outside labels so that the extracted surface best explains those measurements. In that setting, the volumetric labeling is a latent representation of an already well-defined reconstruction target.

In our case, the input is instead a defective triangle mesh whose volumetric meaning is itself ambiguous. Holes, self-intersections, and near-zero-thickness structures do not merely make the correct labeling unknown; they make the inside-outside relation underdetermined from the surface alone. The task is therefore not to fit observations with a surface, but to select a conservative solid interpretation that restores a globally consistent interior-exterior partition.

This difference in problem structure directly changes the formulation. Rather than using symmetric data fidelity terms derived from observations, we preserve only proxy-supported interior evidence through one-sided anchoring, while leaving the remaining space free to be optimized. Likewise, the pairwise term is not a generic smoothness prior: it distinguishes proxy-supported interfaces from unsupported newly introduced ones, and penalizes the latter more strongly to favor compact completion. As a result, graph cuts serve here not as an observation-fitting backend, but as the optimizer of an asymmetric volumetric interpretation problem whose output is a strictly watertight solid induced by the final cell partition.

\section{Method}
\begin{figure}
    \centering
    \includegraphics[width=1\linewidth]{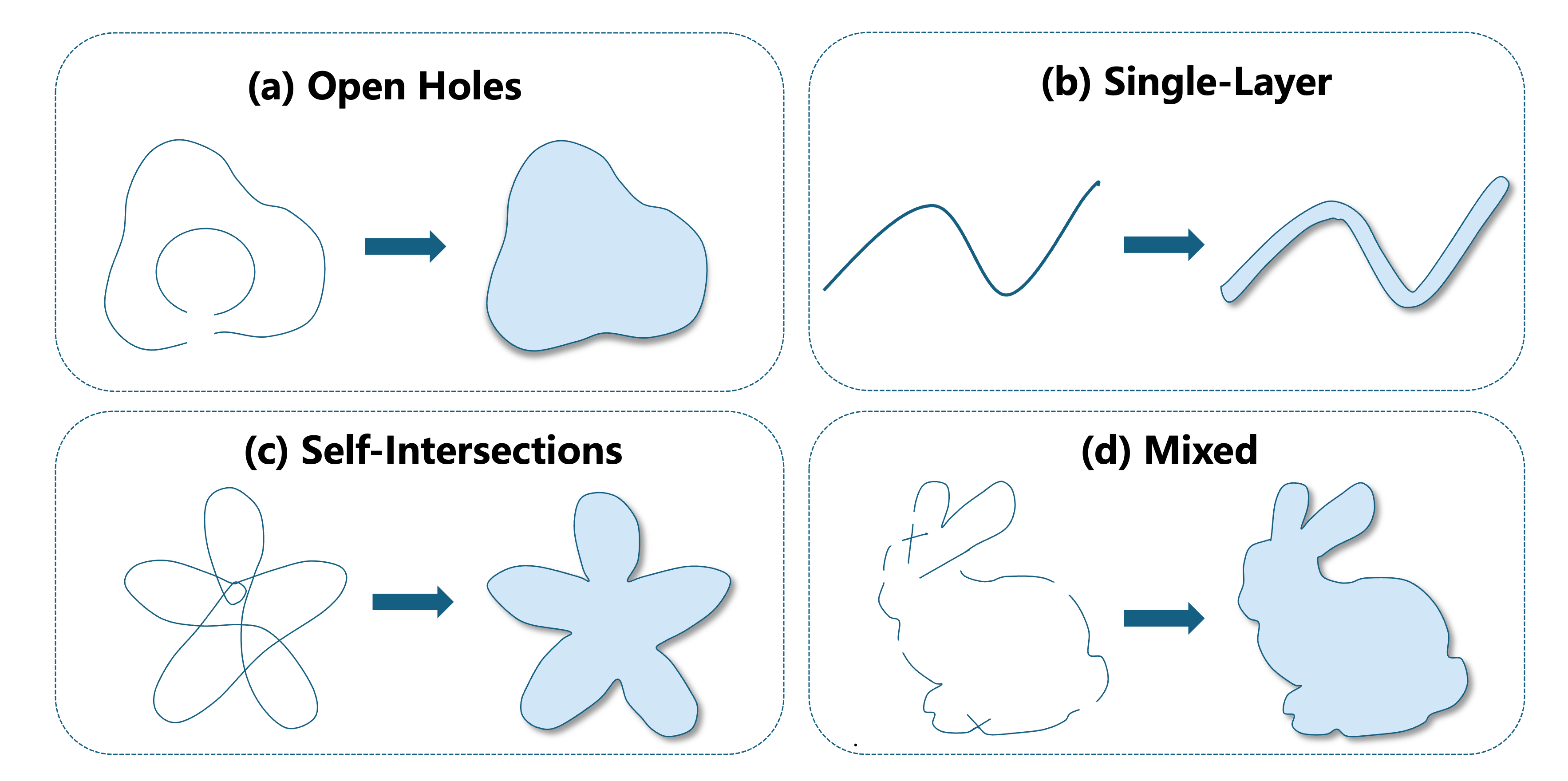}
    \caption{Typical defect patterns that make watertight conversion ambiguous. Open holes, single-layer structures, self-intersections, and their mixtures break a unique interior–exterior interpretation of the input. The filled shapes illustrate possible watertight solid completions.}
    \label{fig:method}
\end{figure}
Before introducing the formulation, we first illustrate why defective meshes are difficult to convert into watertight solids. As shown in Fig.~\ref{fig:method}, common defects such as open holes, single-layer structures, self-intersections, and their mixtures do not merely break manifoldness locally; more fundamentally, they destroy a unique interior--exterior interpretation of the input surface. In these cases, the task is no longer to repair a surface combinatorially, but to select a conservative and globally consistent solid interpretation. This observation motivates our formulation of watertight remeshing as volumetric partitioning rather than local surface repair.

Our design is therefore guided by three principles. First, when the surface no longer defines a unique volume, the algorithm should reason over volumetric interpretations rather than local surface edits. Second, the available geometric evidence is asymmetric: interior support induced by a conservative thickened proxy is trustworthy, whereas the remaining space is ambiguous and should remain optimizable. Third, among all feasible solid interpretations, the preferred one should introduce as little unsupported interface as possible, so that missing regions are completed compactly rather than by creating fragmented or multi-layer boundaries.

These principles directly motivate the three ingredients of our formulation: proxy thickening, one-sided interior anchoring, and fill-aware interface regularization. Based on them, CelloCut embeds the input into a tetrahedral partition of space, assigns each cell a binary interior-exterior label, and optimizes this labeling under constraints that preserve proxy-supported evidence while resolving ambiguity elsewhere. The final surface is then extracted as the interface between interior and exterior cells, yielding a strictly watertight mesh by construction. The detailed steps of our method are summarized in Algorithm~\ref{alg:cellocut}, and the overall pipeline is illustrated in Fig.~\ref{fig:overview_pipeline}.

\begin{algorithm}[!t]
\caption{CelloCut Watertight Remeshing Pipeline}
\label{alg:cellocut}
\begin{algorithmic}[1]
\Require Input mesh $M$, Offset $\epsilon$, Weight $\lambda_{\text{fill}}$
\Ensure Watertight mesh $S_{\text{out}}$

\State \textbf{Stage 1: Discretization (Sec.\ref{sec:thicken_proxy} and Sec.\ref{sec:volumetric_discr})}
\State $\phi(x) \leftarrow \text{ComputeUDF}(M) - \epsilon$ \Comment{Thickening field (Eq. 1)}
\State $S_{\text{thick}} \leftarrow \text{MarchingCubes}(\phi)$
\State $S_{\text{sim}} \leftarrow \text{Decimate}(S_{\text{thick}})$ \Comment{Obtain simplified proxy}
\State $\mathcal{T} \leftarrow \text{DelaunayTetrahedralization}(\text{Vertices}(S_{\text{sim}}))$ \Comment{Volumetric domain $DT(V)$}
\State Compute initial labels $L^*(c) \in \{0, 1\}$ via $\phi(\text{centroid}(c))$

\State \textbf{Stage 2: Graph Construction (Sec.\ref{sec:global-opt})}
\State Initialize graph $G=(\mathcal{V}, \mathcal{E})$ with nodes from cells in $\mathcal{T}$
\For{each cell $c \in \mathcal{T}$} \Comment{Unary Term Construction}
    \If{$L^*(c)=0$} 
        \State Set source edge capacity $\infty$ \Comment{preserve interior}
    \EndIf
\EndFor

\For{each adjacent pair $(c_i, c_j)$ with face $f_{ij}$} \Comment{Binary Term}
    \State $cost \leftarrow \text{Area}(f_{ij})$
    \If{$L^*(c_i) = L^*(c_j)$} \Comment{Penalize forming new boundaries}
        \State $cost \leftarrow cost \cdot \lambda_{\text{fill}}$
    \EndIf
    \State Add edge $(c_i, c_j)$ to $\mathcal{E}$ with capacity $cost$
\EndFor

\State \textbf{Stage 3: Optimization \& Extraction (Sec.\ref{sec:surface_extra})}
\State $L^{\text{opt}} \leftarrow \text{MaxFlowMinCut}(G)$
\State $\mathcal{F} \leftarrow \{ f_{ij} \mid c_i, c_j \text{ adjacent } \land L^{\text{opt}}(c_i) \neq L^{\text{opt}}(c_j) \}$
\State $S_{\text{out}} \leftarrow \text{MarchingCubes}(\text{SDF}(\mathcal{F}))$ \Comment{Refined extraction}

\State \Return $S_{\text{out}}$
\end{algorithmic}
\end{algorithm}

\subsection{Thickened Proxy Surface Generation}\label{sec:thicken_proxy}
To robustly handle input meshes with extremely thin structures, open boundaries, or locally missing surfaces, we avoid operating directly on the defective geometry. Instead, we construct a thickened proxy surface that provides a spatial separation between the interior and exterior.

We first compute the Unsigned Distance Field (UDF) of the input mesh $M$ on a regular grid. To resolve ambiguities in single-layer regions, we introduce a thickening offset $\epsilon$ and define a scalar field: $\phi(x) = UDF_M(x) - \epsilon$,
which provides stable spatial support for structures that originally have zero or near-zero thickness. Meanwhile, the sign of $\phi(x)$ induces a consistent interior–exterior reference with respect to this thickened surface.


We extract the isosurface $\phi(x)=0$ using Marching Cubes, yielding a thickened proxy surface $S_{\text{thick}}$.
We then apply mesh decimation to $S_{\text{thick}}$ to obtain a simplified proxy mesh, denoted as $S_{\text{sim}}$.
This simplification significantly reduces the vertex count for efficient volumetric discretization while preserving the global topological structure of the thickened proxy surface.
Note that $S_{\text{sim}}$ serves as a conservative geometric prior defining the feasible solution space, rather than the final output.

\subsection{Volumetric Discretization and Geometric Prior}\label{sec:volumetric_discr}

Distinguishing interior from exterior regions is critical for watertight remeshing, since the output surface is extracted as the interface between volumetric labels, yielding a closed watertight boundary by construction.
More importantly, embedding defective input geometry into a volumetric domain enables global interior--exterior reasoning, mitigating local ambiguities caused by holes, self-intersections, and missing regions that cannot be reliably resolved by purely surface-based methods.

To enable a volumetric formulation that supports global interior–exterior reasoning, we require a discretization that adapts to geometric complexity without relying on dense uniform grids.
We therefore utilize unconstrained Delaunay tetrahedralization on the vertices 
$V$ of the simplified proxy mesh $S_{\text{sim}}$ to construct the volumetric solution space, denoted as $DT(V)$, which fills the convex hull of the input geometry.

Unlike uniform voxel grids, which impose a rigid trade-off between memory and resolution, Delaunay tetrahedralization provides inherent geometric adaptivity, allocating tetrahedral cells preferentially in regions of high curvature and complex topology while remaining sparse in flat or empty regions. This enables efficient representation of fine geometric details without excessive memory overhead. In practice, the vertex count is adaptively determined based on the local feature size implied by the decimation step, typically ranging from 100k to 500k for complex scans, balancing sufficient sampling density and numerical stability on defective inputs.

To bootstrap the global optimization, we compute an initial interior–exterior label $L^*(c)$ for each cell $c$ by querying the distance field $\phi(x_c)$ at its centroid and assign the label based on its sign:
\begin{equation}
L^*(c) = 
\begin{cases} 
0  & \phi(x_c) < 0, \\ 
1  & \phi(x_c) \ge 0.
\end{cases}
\end{equation}



The labeling $L^*$ is used only as an initialization derived from the thickened proxy, rather than as the final volumetric partition. Because $\phi$ is induced by the thickened proxy, cells with $\phi(x_c)<0$ lie inside the proxy-supported volume and are therefore treated as conservative interior support rather than as a globally reliable partition of the shape. Nevertheless, since $L^*$ is computed independently for each tetrahedral cell, it may still be locally noisy and globally inconsistent in regions affected by missing data or severe geometric defects. The role of the subsequent optimization is therefore not to preserve this initialization everywhere, but to convert this proxy-induced support into a globally consistent volumetric partition.

\subsection{Global Optimization of Partitioning Energy}\label{sec:global-opt}

Our objective is not to preserve the thickened proxy everywhere, nor to fit the defective input surface as closely as possible. Instead, we seek a conservative watertight partition that preserves only geometrically trustworthy interior support and resolves the remaining ambiguity by introducing the smallest additional boundary necessary to form a valid solid. This perspective is essential: in defective meshes, ambiguity is created primarily by missing or inconsistent boundary evidence, not by the absence of plausible interior support. Accordingly, the optimization should not treat all cells symmetrically.

This directly leads to an asymmetric formulation with two roles. A one-sided unary term preserves proxy-supported interior cells and prevents the optimization from carving away already supported mass. A fill-aware pairwise term regularizes the interior-exterior interface by face area, while assigning a larger penalty to newly introduced unsupported boundaries than to proxy-supported ones. Together, these terms bias the solution toward compact single-shell completion: reliable interior is preserved, while ambiguous regions are absorbed only when doing so reduces unsupported interface.

To instantiate this idea, we perform a binary optimization over the tetrahedral cells of $DT(V)$, where each cell is assigned an interior or exterior label.
Each tetrahedral cell $c$ is assigned a binary label
\begin{equation}
L(c) \in \{0, 1\},
\end{equation}
where $0$ denotes interior and $1$ denotes exterior.
Rather than requiring the full initialization $L^*$ to be preserved, we only enforce the subset of cells initialized as interior to remain interior; all other cells are optimized under pairwise regularization.
Formally, this defines the asymmetric feasible set
\begin{equation}
\mathcal{F}=\{L \mid L(c)\in\{0,1\},\; L^*(c)=0 \Rightarrow L(c)=0,\; \forall c \}.
\end{equation}
Under this feasible set, the optimal volumetric labeling is obtained by
\begin{equation}
L^{opt} = \arg\min_{L \in \mathcal{F}} E(L),
\end{equation}
where the energy is defined as
\begin{equation}
E(L) = \sum_{c} V(c, L(c)) + \sum_{(c_i, c_j) \in \mathcal{N}} D(c_i, c_j) \cdot 1[L(c_i) \neq L(c_j)] .
\end{equation}
Here, $\mathcal{N}$ denotes the set of neighboring tetrahedral cell pairs sharing a common face, and $1[\cdot]$ is the indicator function.

\noindent\textbf{Unary Term: One-Sided Interior Anchoring}

The unary term encodes the only hard constraint in our formulation, but its role is not to enforce agreement with the initialization everywhere. Instead, it preserves only the part of the initialization that is geometrically trustworthy: the proxy-supported interior. This asymmetry reflects the structure of the problem. In defective meshes, missing surfaces and topological defects mainly remove boundary evidence and thus create uncertainty in ambiguous regions, but they do not invalidate interior mass that is already supported by a conservative thickened proxy. As a result, interior evidence is fundamentally more reliable than exterior evidence.

We therefore forbid cells initialized as interior from flipping to exterior, while leaving cells initialized as exterior fully relabelable. In this sense, the optimization is conservative in only one direction: it may absorb ambiguous exterior space into the solid, but it never carves away already supported mass.


Formally, for a candidate label $l \in \{0,1\}$ assigned to cell $c$, the unary term is defined as
\begin{equation}
V(c, l) =
\begin{cases}
+\infty, & L^*(c) = 0 \text{ and } l = 1, \\
0, & \text{otherwise}.
\end{cases}
\end{equation}
Therefore, every feasible labeling must preserve all cells marked as interior by $L^*$, while cells initialized as exterior remain unconstrained by the unary term.
This asymmetric feasible set prevents erosion of the proxy-supported interior and leaves the remaining ambiguity to be resolved by the pairwise regularization.
Consequently, the optimization guarantees a conservative enclosure of the proxy-induced interior support, but it does not guarantee recovery of the author-intended topology in severely ambiguous cases.
Instead, it selects, among feasible labelings, a compact watertight completion favored by the pairwise regularizer.

This asymmetry is crucial. Because proxy-supported interior cells cannot flip to exterior, the optimization cannot open artificial exterior gaps through trusted mass. This prevents erosion and strongly suppresses double-shell-like failure modes, which would otherwise require introducing unnecessary exterior layers near already supported interior regions.

\noindent\textbf{Binary Term: Distinguishing Proxy-Supported and Unsupported Interfaces}

The pairwise term regularizes the interface between interior and exterior cells, but its role is more specific than generic smoothness. In our setting, not all candidate interfaces carry the same meaning. If two neighboring cells lie on different sides of the initialization, their shared face is aligned with a boundary already implied by the thickened proxy and can therefore be regarded as weakly supported by the available geometric evidence. In contrast, if the two cells share the same initial label, introducing a cut between them would create a new interface that is not supported by the proxy. Such an interface should be interpreted as a hypothesis of completion rather than observed structure.

This distinction motivates the weighted pairwise term. We penalize all cuts by face area, but assign a larger weight to newly introduced unsupported interfaces than to proxy-supported ones. As a result, new boundaries are created only when they substantially reduce the total interface needed to obtain a feasible watertight partition.

Accordingly, the pairwise cost is defined as 
\begin{equation}
    D(c_i, c_j) = w_{ij} \cdot \text{area}(f_{ij})
\end{equation}
with weights:
\begin{equation}
w_{ij} =\begin{cases}1, & L^*(c_i) \neq L^*(c_j) , \\
\lambda_{fill}, & L^*(c_i) = L^*(c_j) .
\end{cases}
\end{equation}

The $\lambda_{fill}$ multiplier is the key design element that turns a generic cut objective into a completion prior. Interfaces already present in the initialization correspond to boundaries that are at least weakly supported by the proxy, whereas interfaces introduced between cells with the same initial label are unsupported and must therefore be created by the optimization itself. Assigning a larger penalty to the latter makes such new boundaries expensive unless they substantially reduce the overall cut. This induces a discrete surface-tension effect: missing regions are preferably closed by compact patches, while fragmented, elongated, or multi-layer interfaces become energetically unfavorable. In this sense, $\lambda_{fill}>1$ biases the solution toward conservative, compact, and volumetrically consistent solid completion without requiring an explicit curvature model.

The global optimum $L^{opt}$ is efficiently computed via the standard max-flow min-cut \cite{DBLP:journals/pami/BoykovK04}.

\subsection{Surface Extraction and Geometric Regularization}\label{sec:surface_extra}

Although the optimization in Sec.~\ref{sec:global-opt} already determines the watertight topology of the result, the geometry obtained by directly extracting the interface of the labeled tetrahedral complex remains tied to the intermediate discretization. In other words, the optimized cell partition should be viewed as a topologically reliable proxy: it makes the interior-exterior decision globally and consistently, but it is not yet the final geometric representation.

To decouple topological correctness from geometric quality, we therefore reconstruct a refined Signed Distance Field (SDF) from the optimized labeled complex and extract its zero level set on a high-resolution grid using Marching Cubes. This step does not alter the volumetric decision made in Sec.~\ref{sec:global-opt}; it only refines the geometric realization of the already determined watertight boundary.

This two-stage strategy leverages tetrahedral discretization for robust global topology while relying on grid-based extraction for high-quality output geometry.

\section{Experiments}
\subsection{Overview}

To standardize evaluation for watertight remeshing, we introduce a comprehensive suite consisting of large-scale robustness testing on ModelNet10, together with two newly curated, highly challenging benchmarks---CelloScan (geometric fidelity on raw scans) and CelloFill (visual quality in hole filling). We compare our method against six representative baselines: MeshFix\cite{DBLP:journals/vc/Attene10}, ManifoldPlus\cite{huang2020manifoldplus}, VolumeMesher\cite{DBLP:journals/tog/DiazziA21}, fTetWild\cite{DBLP:journals/tog/HuSWZP20}, Dora\cite{DBLP:conf/cvpr/ChenZLLLLLLFT25}, and Craftsman\cite{DBLP:conf/cvpr/LiLYCLCTL25}. We report both geometric and perceptual metrics, with all outputs converted to OBJ format for fair comparison. We do not separately include TetWild\cite{Hu:2018:TMW:3197517.3201353}, since fTetWild has been reported to achieve similar performance while being substantially faster, and thus serves as a representative baseline for this class of methods.


\subsection{Evaluation Datasets and Metrics}
\subsubsection{Robustness Evaluation on ModelNet10}

Following ManifoldPlus~\cite{huang2020manifoldplus}, we evaluate watertight robustness on ModelNet10 dataset. 
For each output, we count the number of boundary edges (Boundary), non-manifold edges (NM Edges), non-manifold vertices (NM Vertices), and triangle inversions (Tri.Inversion). 

\subsubsection{CelloScan}

CelloScan is constructed from topologically challenging 3D meshes collected from Objaverse, the Tencent Hunyuan3D Watertight Conversion Challenge, and additional in-the-wild sources. These inputs exhibit severe non-watertightness, missing regions, self-intersections, and complex topology, making them unsuitable for evaluation under standard input-reconstruction protocols.

\noindent\textbf{Rethinking the Ground Truth.}
Directly treating the defective input mesh as ground truth is problematic for watertight conversion, because the task is not to reproduce the raw surface as-is, but to recover a valid enclosing boundary. In many real inputs, the mesh contains topological defects, internal fragments, or ambiguous thin structures that are incompatible with a single consistent solid interpretation. Under such cases, input-based geometric metrics become misaligned with the task: they reward methods for staying close to defective geometry, even when doing so yields volumetrically implausible or non-manifold outputs.

We therefore evaluate against the object's \emph{visible outer surface}, approximated using a virtual scanning (ray-casting) procedure. By aggregating externally observed geometry from multiple viewpoints, this protocol captures the effective enclosing boundary while suppressing interior clutter and defect-specific artifacts. The resulting reference is reproducible and applied uniformly to all methods, providing a more task-aligned benchmark for watertight conversion than direct comparison to the raw defective input.

Importantly, this protocol does not encode our method’s output or assume a particular reconstruction strategy. It only captures geometry that is externally observable from the object boundary and is applied identically to all methods. In this sense, the reference favors neither our volumetric formulation nor any specific completion mechanism; rather, it evaluates whether a method recovers a valid enclosing boundary that is consistent with the object’s visible outer surface.

\noindent\textbf{Geometric Evaluation Metrics.} Reconstruction quality is evaluated by sampling points from the output mesh and comparing them against this Outer Surface GT point cloud. We report Chamfer Distance (CD), Hausdorff Distance (HD), Absolute Normal Consistency (ANC), and F-score (F1), where F1 is computed under a fixed distance threshold. This protocol ensures that metrics measure fidelity to the object's restored physical boundary, rather than adherence to a defective input or noisy reference topology.

\subsubsection{CelloFill}

CelloFill is a synthetic benchmark derived from watertight 3D models from the Google Scanned Objects (GSO) dataset. 
To simulate realistic scanning defects, we algorithmically introduce holes into the original meshes, creating inputs that resemble typical sensor artifacts: holes at grazing-angle surfaces, elongated gaps caused by motion or anisotropic sampling, and small stripe-like dropouts akin to scan-line failures. 
Each input mesh contains $6$--$12$ holes of varying size and shape, visually significant yet preserving the overall topology. 
The original watertight mesh serves as ground truth, allowing controlled evaluation of the algorithm’s ability to reconstruct missing geometry rather than replicate the defective input.

\noindent\textbf{Evaluation Metrics.} 
Given the significant scanning artifacts modeled in CelloFill, our goal is plausible geometry reconstruction rather than strict pixel replication. 
We render untextured RGB images (clay renders) from six canonical views under identical lighting, materials, and camera settings for all methods to ensure a fair comparison. 
We report LPIPS\cite{DBLP:conf/cvpr/ZhangIESW18}, FID\cite{DBLP:conf/nips/HeuselRUNH17}, and CLIP Score\cite{DBLP:conf/icml/RadfordKHRGASAM21} to assess perceptual smoothness and semantic consistency. 
We do not include depth or normal renderings for perceptual evaluation, as these are more directly tied to geometric accuracy, which is already evaluated in Table~\ref{tab:geometric comparison}. 
Pixel-wise metrics (PSNR/SSIM) are excluded, as they are overly sensitive to minor misalignments in large-scale hole filling and do not correlate well with visual plausibility.

These perceptual metrics are intended to evaluate the visual quality of hole completion rather than to replace geometric evaluation. Geometric fidelity is already assessed separately in Table \ref{tab:geometric comparison} under the CelloScan protocol, whereas CelloFill focuses specifically on whether large missing regions are completed in a visually coherent and semantically consistent manner. The two benchmarks are therefore complementary: one measures task-aligned boundary fidelity, and the other measures perceptual quality of completion under controlled missing-geometry corruption.

\subsection{Implementation Details}

Our pipeline is fully deterministic and does not involve learning. 
The unsigned distance field (UDF) is extracted on a $512^3$ grid, and surface extraction is performed via marching cubes with a shell thickness $\epsilon = 1 / 512$. 
Volumetric discretization is obtained using Delaunay tetrahedralization, implemented via CGAL\cite{cgal:eb-26a}, and the volumetric labeling is solved via a graph-cut formulation using PyMaxflow\cite{DBLP:journals/pami/BoykovK04}.  
The thickened proxy surface is simplified using a standard mesh decimation method prior to tetrahedralization, in order to reduce the complexity of volumetric discretization and graph-cut optimization.
Unless otherwise specified, we use $\lambda_{\text{fill}} = 20$ in all experiments. 
For visual evaluation, all renderings are generated using the same rendering pipeline provided by \textit{Trellis}\cite{DBLP:conf/cvpr/XiangLXDWZC0Y25}.

\subsection{Quantitative Results}
\noindent\textbf{ModelNet10 (Robustness)}
Following ManifoldPlus, we evaluate watertight robustness on ModelNet10 \cite{DBLP:conf/cvpr/WuSKYZTX15}. As shown in Table~\ref{tab:Robustness comparison}, CelloCut consistently produces strictly watertight outputs with no open boundaries, non-manifold edges, non-manifold vertices, or failed cases. ManifoldPlus and Dora also achieve zero across all four reported robustness criteria. In contrast, fTetWild fails to reliably produce watertight outputs on this benchmark, with boundary violations on a large number of shapes and also a non-zero number of failures. Craftsman also exhibits non-zero boundary cases, although to a much smaller extent. VolumeMesher frequently violates vertex-manifoldness, resulting in many shapes with non-manifold vertices. Due to this fundamental violation of the strict watertightness requirement, VolumeMesher is excluded from quantitative evaluations on CelloScan and CelloFill. These results demonstrate that our constructive volumetric formulation robustly preserves strict watertight validity across diverse meshes.

\begin{table}[htbp] 
\centering 
\caption{\textbf{Robustness evaluation on ModelNet10.} We report the number of shapes violating strict watertight constraints under four criteria: \textbf{Boundary} (shapes containing one or more open boundary loops or holes), \textbf{NM Edges} (shapes containing one or more non-manifold edges shared by more than two faces), \textbf{NM Vertices} (shapes containing one or more non-manifold vertices), and \textbf{Failure} (shapes for which the method fails to produce a valid output). A strictly watertight surface must achieve zero across all metrics.}

\label{tab:Robustness comparison}
\cellocuttableformat
\begin{tabular}{lcccc}
\toprule
\textbf{Methods} & \textbf{Boundary} & \textbf{NM Edges} & \textbf{NM Vertices} & \textbf{Failure}  \\
\midrule
ManifoldPlus & 0 & 0 & 0 & 0 \\
VolumeMesher & 0 & 0 & 2420 & 0 \\
Dora         & 0 & 0 & 0 & 0  \\
fTetWild     &4872 &0 &0 &20\\
Craftsman    & 40  & 0  & 0  & 0  \\
Ours         & 0 & 0 & 0 & 0  \\
\bottomrule
\end{tabular}
\end{table}

\noindent\textbf{CelloScan (Geometric Fidelity)}
Table \ref{tab:geometric comparison} reports geometric performance on CelloScan using outer surface ground truth from virtual scanning. CelloCut achieves the best overall results in CD, ANC, and F1@0.01, indicating more accurate recovery of visible boundaries. ManifoldPlus and Dora obtain slightly lower HD but underperform on CD and F1, suggesting sensitivity to local inconsistencies. fTetWild yields weaker CD, HD, ANC, and F1 on CelloScan, and also shows a small number of failures. This may be because it mainly targets robust tetrahedral meshing within a geometric tolerance, whereas our benchmark emphasizes watertight completion under larger missing regions. VolumeMesher achieves reasonable accuracy on some samples but shows a high failure rate, as it does not explicitly enforce watertight or manifold constraints. By incorporating additional structural constraints within a volumetric framework, CelloCut produces more robust and consistently valid reconstructions.
\begin{table}[htbp]
\centering
\caption{Geometric metrics on our newly introduced CelloScan dataset using virtual-scanned outer surfaces as ground truth, showing our framework consistently outperforms existing mesh repair, tetrahedral remeshing, and grid-based methods.}
\label{tab:geometric comparison}
\cellocuttableformat
\setlength{\tabcolsep}{6pt}
\begin{tabular}{lccccc}
\toprule
\textbf{Methods} &\textbf{Failure$\downarrow$} & \textbf{CD$\downarrow$} & \textbf{HD$\downarrow$} & \textbf{ANC$\uparrow$} & \textbf{F1@0.01$\uparrow$}  \\
\midrule
MeshFix          &  40 & 0.202739 & 0.902212 & 0.5638 & 3.48  \\
ManifoldPlus     &  0 & 0.000074 & \textbf{0.083855} & 0.9363 & 95.65 \\

fTetWild         & 4 & 0.000808 & 0.134098 & 0.9057 & 84.38\\
Dora             &  0 & 0.000623 & 0.084281 & 0.9402 & 92.94  \\
Craftsman        &  2 & 0.000463 & 0.103374 & 0.9327 & 90.48  \\
Ours             &  0 & \textbf{0.000048} & 0.089819 & \textbf{0.9452} & \textbf{96.96}  \\
\bottomrule
\end{tabular}
\end{table}

\noindent\textbf{CelloFill (Visual Quality). }
Table~\ref{tab:visual comparison} summarizes perceptual results on CelloFill under large missing regions. Perceptual metrics are computed on valid watertight outputs only. CelloCut achieves the lowest FID and competitive LPIPS, demonstrating strong visual coherence. MeshFix attains favorable perceptual scores on a limited subset of successful cases, reflecting its ability to aggressively complete missing geometry; however, under this protocol, it overwhelmingly fails to produce watertight outputs, resulting in a substantially high failure rate. Other baseline methods are generally less effective at recovering large missing regions, which naturally leads to lower visual similarity to the watertight ground truth. In contrast, CelloCut consistently recovers compact, watertight solids while maintaining high visual quality.

\begin{table}[htbp]
\centering
\caption{Visual metrics (FID, LPIPS, and CLIP) under large missing regions evaluated on CelloFill dataset, showing our framework consistently outperforms existing shape completion and watertight remeshing methods.}
\label{tab:visual comparison}
\cellocuttableformat
\begin{tabular}{lccc}
\toprule
\textbf{methods}  &\textbf{FID$\downarrow$} & \textbf{LPIPS$\downarrow$} & \textbf{CLIP$\uparrow$}   \\
\midrule
MeshFix            & 29.4505 & 0.0513 & \textbf{98.66}  \\
ManifoldPlus   & 69.7035 & 0.1223 & 93.61  \\
fTetWild & 39.9217 & 0.0555 & 97.09 \\
Dora               & 37.1267 & 0.0645 & 96.99  \\
Craftsman           & 30.7946 & 0.0504 & 97.88  \\
Ours             & \textbf{20.1733} & \textbf{0.0465} & 98.42  \\

\bottomrule
\end{tabular}
\end{table}

\subsection{Qualitative Results}
\begin{figure*}

  \includegraphics[width=\linewidth]{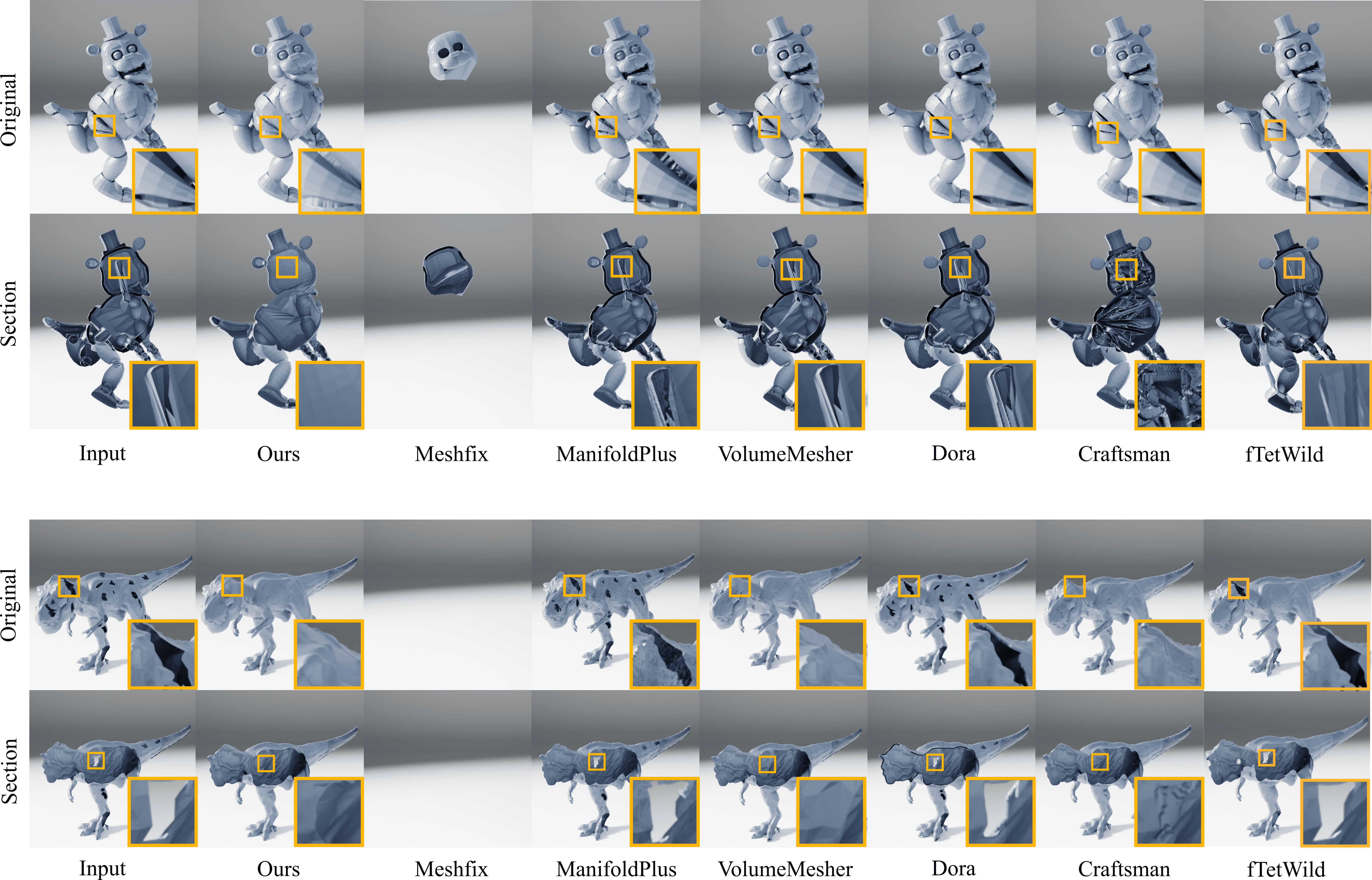}

  \caption{\textit{Qualitative comparison of watertight remeshing.}
Odd columns show surface reconstructions and even columns show longitudinal sections, revealing differences in volumetric consistency, hole sealing, and internal structure.}

  \label{fig: Qualitative results} 
\end{figure*}

\noindent\textbf{Qualitative Comparisons.} 
Qualitative evaluation of watertight remeshing results is shown in Fig.~\ref{fig: Qualitative results}, comparing our method with
\textbf{MeshFix}, \textbf{ManifoldPlus}, \textbf{VolumeMesher}, \textbf{Dora}, \textbf{Craftsman} and \textbf{fTetWild} on two representative models: \textit{bad\_doll} and \textit{dinosaur}. 
The examples illustrate common topological and geometric challenges in real-world scanned data, including large semantic openings, self-intersections, and non-manifold configurations.

For the \textit{bad\_doll} model, the object is globally recognizable but severely corrupted by large semantic openings around articulated regions, together with articulated gaps and self-intersections, which cannot be resolved by simple geometric hole filling.
MeshFix preserves only a small connected component and discards most of the shape.
ManifoldPlus produces a watertight surface but introduces sharp spikes in repaired regions.
VolumeMesher preserves local geometry but fails to close the large semantic opening on the back, resulting in a non-watertight output.
Dora generates a double-shell structure with low geometric fidelity, fTetWild fails to close the larger openings, and Craftsman introduces noticeable grid-like artifacts.
In contrast, our method successfully closes both large and small openings and reconstructs a clean, single-shell watertight surface without internal fragments.
For the \textit{dinosaur} model, the mesh has many small holes and self-intersections. VolumeMesher, Craftsman, and our method fill most gaps, but VolumeMesher violates vertex-manifoldness, Craftsman produces stair-step artifacts, and fTetWild remains less effective on the larger missing regions. ManifoldPlus and Dora fail similarly as before, and MeshFix cannot produce a valid mesh. In contrast, our approach robustly resolves holes and self-intersections, producing a smooth, watertight mesh with consistent interior–exterior definition suitable for downstream tasks.

\section{Discussion}

We further analyze CelloCut from three perspectives: the sensitivity of the two key parameters, the necessity of the main structural components, and the current limitations of the framework. In particular, we study the effect of the thickening offset $\varepsilon$, the filling weight $\lambda_{\text{fill}}$, and the two core design choices that make the formulation work in practice, namely surface thickening and the one-sided unary term. Additional implementation-oriented analyses are provided in the supplementary material, including the geometric effects of mesh simplification and tetrahedralization, sensitivity to the simplification ratio, runtime comparisons, and stage-wise runtime breakdowns. Additional qualitative comparisons are also included in the supplementary material.

\begin{figure}[t]
  \centering
  \begin{subfigure}{0.32\columnwidth}
    \centering
    \includegraphics[width=\linewidth]{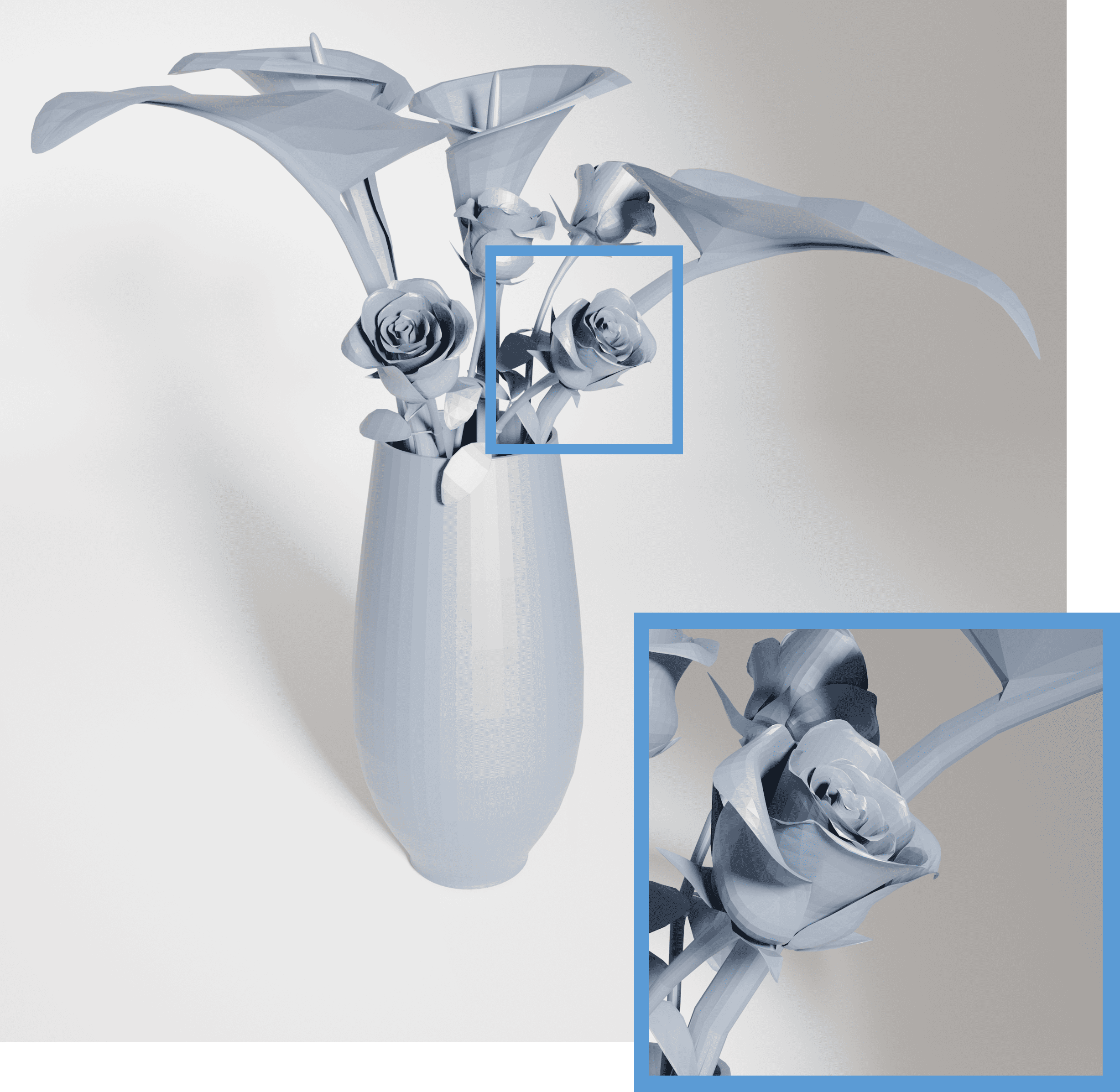}
    \caption{Raw}
    \label{fig:raw}
  \end{subfigure}
  \hfill 
  \begin{subfigure}{0.32\columnwidth}
    \centering
    \includegraphics[width=\linewidth]{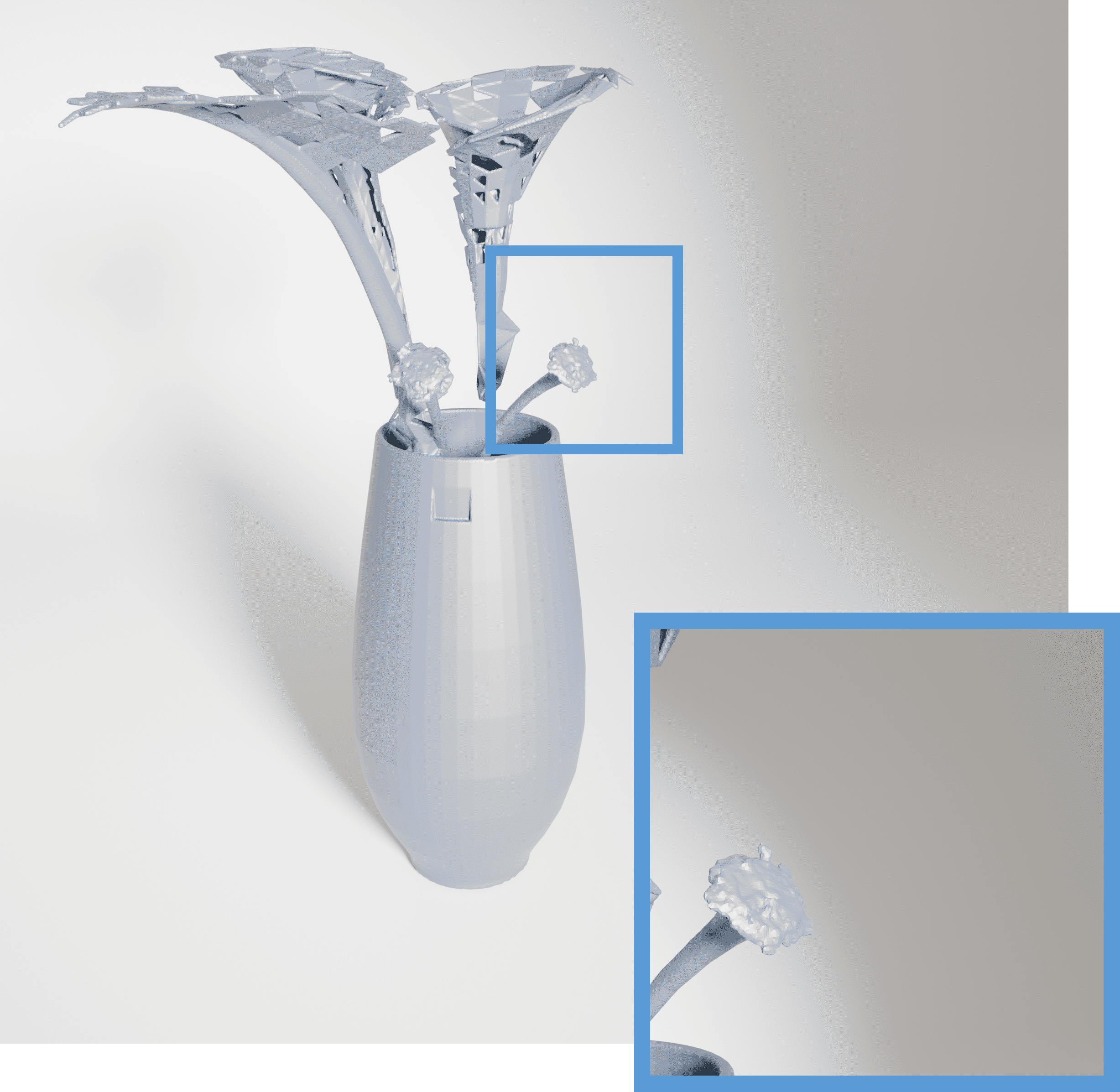}
    \caption{w/o thickening}
    \label{fig:no_thick}
  \end{subfigure}
  \hfill
  \begin{subfigure}{0.32\columnwidth}
    \centering
    \includegraphics[width=\linewidth]{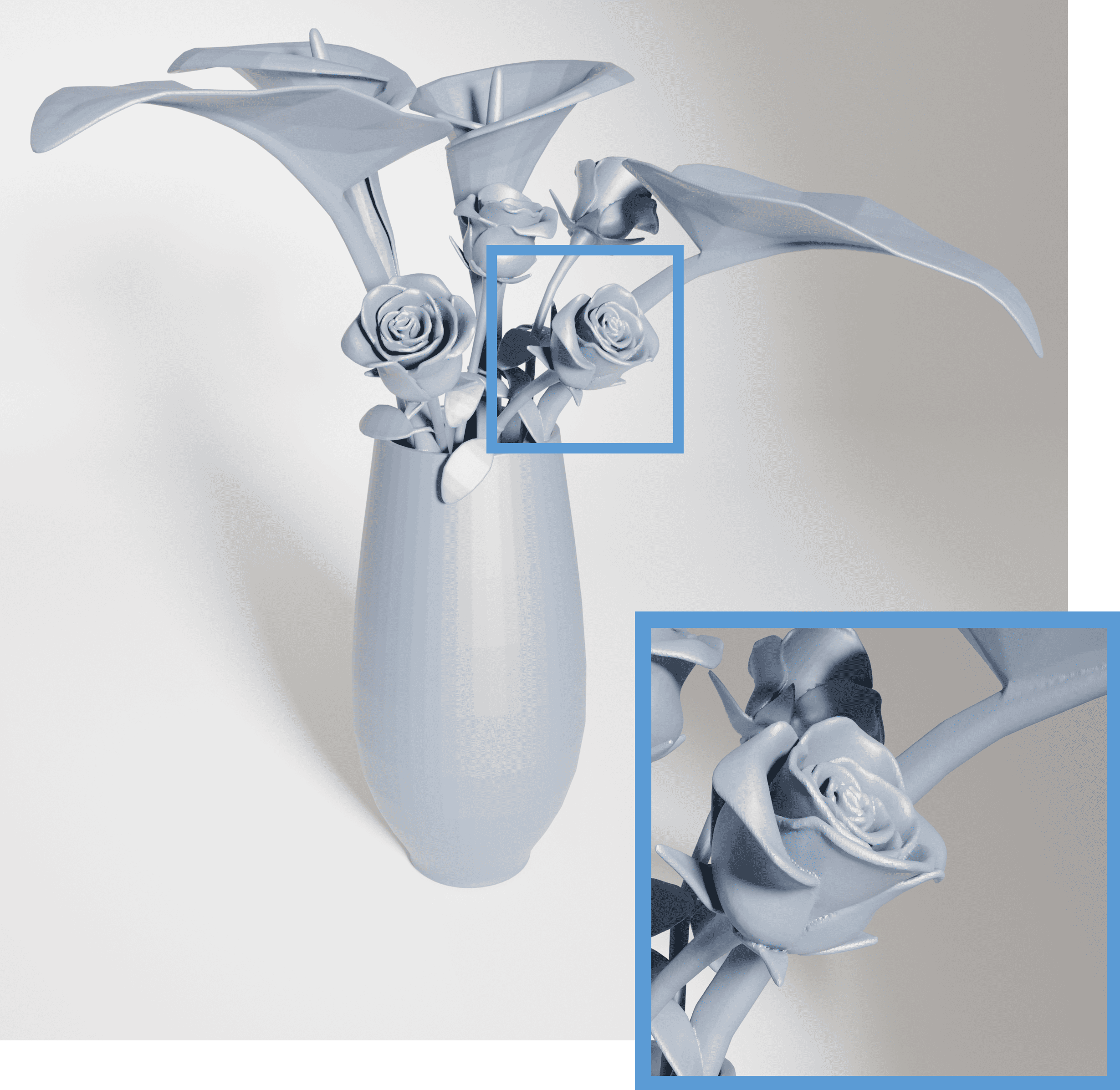}
    \caption{w/ thickening}
    \label{fig:thick}
  \end{subfigure}
  \caption{Effect of Surface Thickening. Removing the thickening step causes thin structures to collapse and introduces topological instability (b), whereas our full method successfully recovers a robust solid (c).
}

\label{fig:thickening_comparison}
\end{figure}
\subsection{Parameter Sensitivity}
\subsubsection{Choice of $\varepsilon$.}
\begin{table}[t]
\centering
\caption{Effect of $\varepsilon$ on reconstruction quality. We report Chamfer-L2 distance, Hausdorff distance, and F1 score at a strict threshold.}
\label{tab:eps}
\cellocuttableformat
\begin{tabular}{c c c c}
\toprule
$\varepsilon$ & Chamfer-L2 $\downarrow$ & Hausdorff $\downarrow$ & F1@0.01 $\uparrow$ \\
\midrule
$0.5/512$ & 0.000406 & 0.119062 & 93.63 \\
$1/512$ & \textbf{0.000048} & \textbf{0.089819} & \textbf{96.96} \\
$1.5/512$ & 0.000068 & 0.089640 & 87.66 \\
$2.0/512$ & 0.000092 & 0.091573 & 75.59 \\
\bottomrule
\end{tabular}
\end{table}

The parameter $\varepsilon$ controls the amount of geometric thickening applied in preprocessing, allowing thin or single-layer structures to induce a stable volumetric proxy for subsequent labeling and surface extraction.
As shown in Table~\ref{tab:eps}, a small $\varepsilon$ ($0.5/512$) fails to close narrow gaps, leading to fragmented surfaces and increased geometric error. 
Conversely, overly large values ($\ge 1.5/512$) oversmooth fine-scale structures, causing a sharp drop in strict F1 scores (F1@0.01), even though global metrics such as Chamfer distance remain relatively stable. 
We therefore choose $\varepsilon = 1/512$ as a balanced setting that minimizes both Chamfer and Hausdorff errors while preserving geometric detail.

\subsubsection{Sensitivity to the Filling Weight $\lambda_{\text{fill}}$}

\begin{figure}[t]
    \centering
    \includegraphics[width=0.62\linewidth,height=0.3\textheight,keepaspectratio]{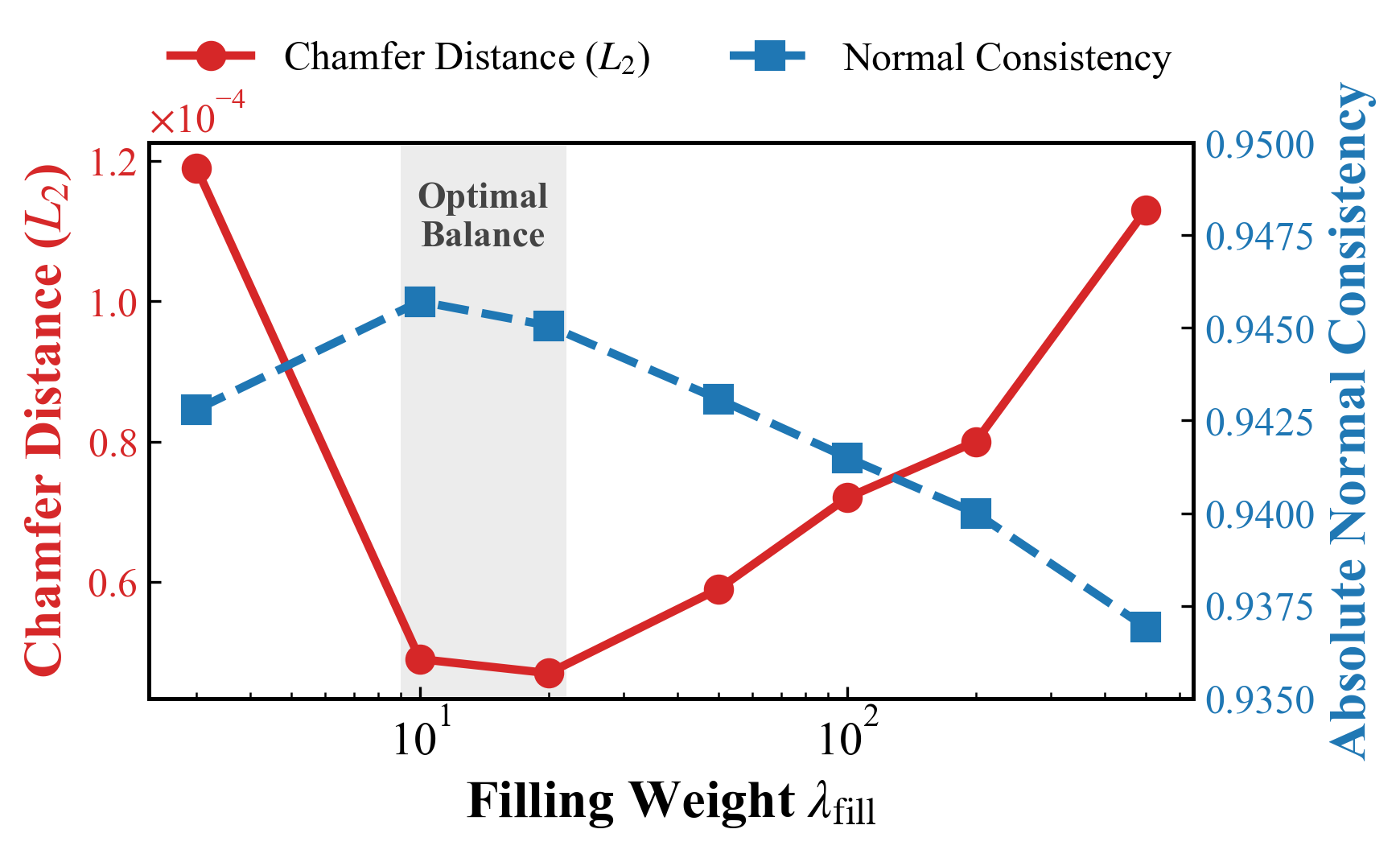}
    \caption{Sensitivity to the filling weight $\lambda_{\text{fill}}$ on the CelloScan dataset.}
    \label{fig:lambda_sensitivity}
\end{figure}

The filling weight $\lambda_{\text{fill}}$ balances aggressive hole filling against maintaining the geometry suggested by the thickened input surface. 
We evaluate its sensitivity on \textbf{CelloScan} by varying $\lambda_{\text{fill}} \in \{3, 10, 20, 50, 100, 200, 500\}$ while keeping all other settings fixed. 
As shown in Figure~\ref{fig:lambda_sensitivity}, small values encourage the creation of new surfaces to fill gaps, which can lead to overfilling and degraded geometric accuracy despite achieving strong topological closure. 
In contrast, very large values heavily penalize new surfaces, causing the optimization to favor minimal changes beyond the initial thickened surface and leaving large holes unfilled. 
Moderate values ($\lambda_{\text{fill}}=10$--$20$) achieve a favorable trade-off, providing low geometric error and consistently high F1 scores. 
These results show that CelloCut behaves predictably across a broad parameter range, and we therefore adopt $\lambda_{\text{fill}}=20$ as the default in all experiments.

\subsection{Ablation Study}
We next isolate the two structural components that are most critical to the formulation itself, namely surface thickening and the one-sided unary constraint, in order to assess whether the observed performance gains come from the full design rather than parameter tuning alone.

\noindent\textbf{Effect of Removing Surface Thickening.}
We evaluate the impact of the thickening offset $\epsilon$.
As shown in Fig.~\ref{fig:thickening_comparison}, disabling thickening causes the reconstruction to fail in thin or incomplete regions.
Quantitatively, this degradation is severe: F1@0.01 drops to $65.2\%$ and the Hausdorff distance increases to $0.45$.
These results indicate that surface thickening is not merely an enhancement, but a prerequisite for resolving intrinsic ambiguities in single-layer geometry.

\noindent\textbf{Effect of Removing the Unary Term.}
We remove the unary term while keeping the pairwise interface regularization unchanged.
In this setting, the optimization collapses to a trivial solution with no interior region, yielding no interior--exterior interface and thus no extractable surface.
Consequently, geometric evaluation is not applicable.
This experiment confirms that the unary term is essential for preventing degenerate solutions and for anchoring the volumetric labeling to the geometric prior.

\subsection{Limitations and Future Works}

Despite its robustness, CelloCut still has several limitations.
Although parts of the pipeline are GPU-accelerated, the overall runtime is dominated by global graph-cut optimization, whose sequential nature limits parallelization and real-time performance on high-resolution inputs. Detailed runtime comparisons and stage-wise runtime breakdowns are provided in the supplementary material.
Moreover, the output surface density is tied to the volumetric discretization, which can cause unnecessarily dense tessellation in geometrically simple regions.
Developing adaptive discretization schemes and parallel or approximate solvers could substantially improve scalability and efficiency.

Beyond efficiency, an important future direction is to integrate CelloCut’s deterministic topological guarantees with learned semantic priors, potentially enabling more informed volumetric labeling in severely ambiguous or incomplete regions.
In particular, recent work on solid--shell labeling for discrete surfaces \cite{DBLP:conf/siggrapha/0003MGRZPL25} shows that many in-the-wild meshes do not admit a single uniform solid interpretation, but instead contain a mixture of solid and shell elements.
This suggests an important limitation of our current formulation, which assumes uniform solidification. Extending CelloCut toward semantics-aware volumetric reasoning, potentially with sparse user guidance, is therefore a promising direction for handling such mixed-structure inputs.

\section{Conclusion} 
We presented CelloCut, a framework that reformulates watertight remeshing as a constrained volumetric partitioning problem.
By shifting from surface-based inference to global tetrahedral labeling, our method guarantees strictly watertight outputs by construction and robustly resolves fundamental ambiguities such as single-layer structures and large missing regions.
Extensive experiments demonstrate that CelloCut consistently produces globally consistent solids in cases where state-of-the-art methods fail.

\bibliographystyle{ACM-Reference-Format}
\bibliography{reference}

\clearpage
\section*{Supplementary Material}
\addcontentsline{toc}{section}{Supplementary Material}
\section{Additional Implementation Details}
\subsection{Practical Notes}

Several implementation details are worth noting.
First, tetrahedralization is performed on a simplified thickened proxy to reduce graph complexity without affecting volumetric topology.
Second, one-sided unary constraints ensure that high-confidence interior regions are strictly preserved during optimization.
Finally, rather than directly using the boundary of the tetrahedral mesh, we first extract all faces shared by adjacent tetrahedra with opposite labels, and then convert this extracted interface into a signed distance field, from which the final surface is reconstructed using marching cubes.
This decoupled extraction step separates topological correctness from geometric refinement, allowing the optimized cell partition to guarantee watertightness while the final surface is reconstructed from a smoother volumetric representation.

\section{Extended Analyses}

This supplementary material complements the main paper with implementation-oriented and extended empirical analyses that are not central to the main narrative. Since parameter sensitivity (\(\varepsilon\) and \(\lambda_{\text{fill}}\)) and the core ablations are already presented in the main text, we focus here on three additional aspects: the geometric effects of intermediate discretization, runtime characteristics, and the impact of the mesh simplification ratio. These analyses help clarify how CelloCut behaves in practice beyond the main experimental conclusions.

\subsection{Geometric Effects of Mesh Simplification and Tetrahedralization}
To verify that the intermediate discretization steps do not dominate the final geometric behavior, we measure the deviation introduced by mesh simplification and tetrahedralization before volumetric optimization. Each intermediate representation is compared against the thickened proxy surface, which serves as the geometry-preserving reference prior to graph-cut optimization.

\begin{table}[htbp] %
    \centering
    \caption{Geometric deviation introduced by mesh simplification and tetrahedralization, measured against the thickened proxy surface.}
    \label{tab:geo_simplification_tet}
    \cellocuttableformat
    \begin{tabular}{lcccc}
        \toprule
        \textbf{Comparison} & \textbf{CD} & \textbf{HD} & \textbf{ANC} & \textbf{F1@0.01} \\
        \midrule
        Thi vs. Sim & 0.000030 & 0.020267 & 0.9491 & 96.59\\
        Thi vs. Tet & 0.000031 & 0.034036 & 0.9047 & 96.49 \\
        Sim vs. Tet & 0.000031 & 0.029300 &0.9053 & 96.23\\
        \bottomrule
    \end{tabular}
\end{table}

As shown in Table~\ref{tab:geo_simplification_tet}, mesh simplification preserves the overall geometry with negligible deviation, yielding a Chamfer-L2 distance of $3.0\times10^{-5}$ and an F1 score of 96.59 at a threshold of 0.01. 
Tetrahedralization introduces slightly larger local deviations, as the smooth surface is represented by flat tetrahedral faces, causing minor differences in geometry and normals.
These deviations are small compared to the geometric changes introduced by topological repair and hole filling, demonstrating that our discretization steps introduce limited bias into the final reconstruction.

\subsection{Runtime Discussion}

To complement the discussion of computational limitations in the main paper, we report the average runtime of different methods on the CelloScan benchmark. The goal here is not only to provide a practical efficiency reference, but also to clarify where the computational cost of CelloCut arises in comparison with representative baselines.

All experiments are conducted on a single machine equipped with one NVIDIA RTX PRO 6000 GPU (96GB) and a CPU with 22 vCPUs (Intel(R) Xeon(R) Platinum 8470Q).
Reported numbers correspond to the average processing time per model, measured in seconds.

\begin{table}[htbp]
    \centering
    \caption{Average runtime (in seconds) on the CelloScan benchmark.}
    \label{tab:runtime}
    \cellocuttableformat
    \begin{tabular}{lc}
        \toprule
        \textbf{Method} & \textbf{Time (s)} \\
        \midrule
        Ours        & 74.62 \\
        Dora        & 19.55 \\
        Craftsman   & 110.35 \\
        ManifoldPlus & 10.95 \\
        \bottomrule
    \end{tabular}
\end{table}

\begin{table}[htbp]
    \centering
    \caption{Average runtime breakdown of CelloCut on the CelloScan benchmark. We report the percentage of total runtime spent in each major stage.}
    \label{tab:runtime_breakdown}
    \cellocuttableformat
    \begin{tabular}{lc}
        \toprule
        \textbf{Stage} & \textbf{Time Ratio (\%)} \\
        \midrule
        Graph-cut optimization & 64.60 \\
        Initial labeling + adjacency construction & 20.08 \\
        Final extraction & 10.20 \\
        Tetrahedralization & 3.39 \\
        Mesh decimation & 1.53 \\
        UDF computation & 0.10 \\
        Marching Cubes & 0.10 \\
        \bottomrule
    \end{tabular}
\end{table}

ManifoldPlus and Dora are faster than CelloCut because they avoid the global volumetric optimization used in our method. 
By contrast, CelloCut requires additional computation for volumetric discretization and graph-cut optimization in order to explicitly enforce strict watertightness and manifoldness. 
Craftsman is the slowest among the compared methods, mainly due to its dense mesh-to-SDF conversion and rendering-based visibility estimation.

We do not report runtime results for MeshFix, fTetWild, or VolumeMesher on CelloScan, because these methods frequently time out, fail to produce valid outputs, or both on the most challenging inputs, making average runtime comparisons incomplete and potentially misleading.

To better understand where the computational cost of CelloCut arises, we further profile the runtime distribution of our pipeline on CelloScan. As shown in Table~\ref{tab:runtime_breakdown}, graph-cut optimization dominates the total runtime, accounting for 64.60\% on average. The second largest component is the combined stage of initial labeling and adjacency construction (20.08\%), which includes computing tetrahedron centroids, querying volumetric values for all cells, and constructing face adjacencies with hashing-based lookup and area evaluation. Final extraction accounts for 10.20\% of the runtime, while tetrahedralization and mesh decimation contribute 3.39\% and 1.53\%, respectively. By comparison, UDF computation and marching cubes each account for only about 0.10\% of the total runtime.

These results show that the runtime of CelloCut is dominated by global optimization, with a substantial additional cost from cell-level initialization and adjacency construction. In particular, both graph-cut optimization and the preparation of tetrahedral cell relationships become more expensive as the volumetric discretization grows, which further motivates the mesh simplification strategy adopted in our pipeline.

\subsection{Effect of Mesh Simplification Ratio}

Our pipeline includes a mesh simplification step applied to the thickened proxy surface prior to tetrahedralization.
This step is introduced to reduce the complexity of the volumetric discretization and the subsequent graph-cut optimization by removing redundant surface triangles, while preserving the overall geometry and topology.

\begin{table}[htbp]
\centering
\caption{Effect of mesh simplification ratio on geometric accuracy.}
\label{tab:simplification_ablation}
\cellocuttableformat
\begin{tabular}{lcccc}
\toprule
\textbf{Simplification Ratio} & \textbf{CD$\downarrow$} & \textbf{HD$\downarrow$} & \textbf{ANC$\uparrow$} & \textbf{F1@0.01$\uparrow$}  \\
\midrule
0.8       &  0.000049 & 0.088333 & 0.9466 &  96.63 \\
0.9       &  0.000049 & 0.087676 & 0.9468 &  96.67 \\
0.95      & 0.000048 & 0.089819 & 0.9452 & 96.96\\
0.98      & 0.000050 & 0.091052 & 0.9358 & 96.58\\
\bottomrule
\end{tabular}
\end{table}

To study how mesh simplification affects the pipeline itself, we evaluate several proxy decimation settings before tetrahedralization. Here, the simplification ratio denotes the \emph{fraction of faces removed} from the thickened proxy surface. We test ratios of 0.8, 0.9, 0.95, and 0.98, corresponding to removing 80\%, 90\%, 95\%, and 98\% of the original faces, respectively.

For geometric analysis, we compare the final reconstructed surfaces produced under different simplification settings against the output of the corresponding unsimplified pipeline. This experiment is intended as a relative robustness study of the internal discretization strategy, rather than as a replacement for the dataset-level evaluation protocol used in the main paper. We report Chamfer distance, Hausdorff distance, absolute normal consistency, and F1@0.01. We focus here on geometric sensitivity; the computational motivation for simplification is discussed separately in the runtime analysis above.
The quantitative results are reported in Table~\ref{tab:simplification_ablation}.

Overall, we observe that moderate to aggressive simplification has a limited effect on geometric accuracy.
Chamfer Distance and F1 score remain highly stable across different ratios, while normal consistency exhibits only minor variations.
Notably, a simplification ratio of 0.95 achieves the best overall trade-off, yielding the lowest Chamfer Distance and the highest F1 score among the tested settings.
Further increasing the ratio to 0.98 leads to a slight degradation in normal consistency and Hausdorff Distance, suggesting that overly aggressive decimation may begin to affect surface fidelity.

From an algorithmic perspective, more aggressive simplification reduces the size of the tetrahedral graph and the number of optimization variables, and is therefore expected to improve runtime. Combined with the small geometric variation observed in Table~\ref{tab:simplification_ablation}, this supports our use of aggressive but controlled simplification in practice.

Based on this trade-off, we adopt a simplification ratio of 0.95 in all experiments.
This setting removes the majority of redundant surface elements while preserving geometric fidelity and numerical stability.
We also observe that even higher simplification ratios may occasionally introduce unstable tetrahedralization or degenerate configurations, and thus we avoid more aggressive decimation.
Overall, this experiment demonstrates that aggressive but controlled mesh simplification is an effective strategy for accelerating the pipeline while maintaining high reconstruction quality.

\section{Additional Qualitative Results}

\begin{figure}[!htbp]
\centering
  \includegraphics[width=0.85\linewidth,height=0.78\textheight,keepaspectratio]{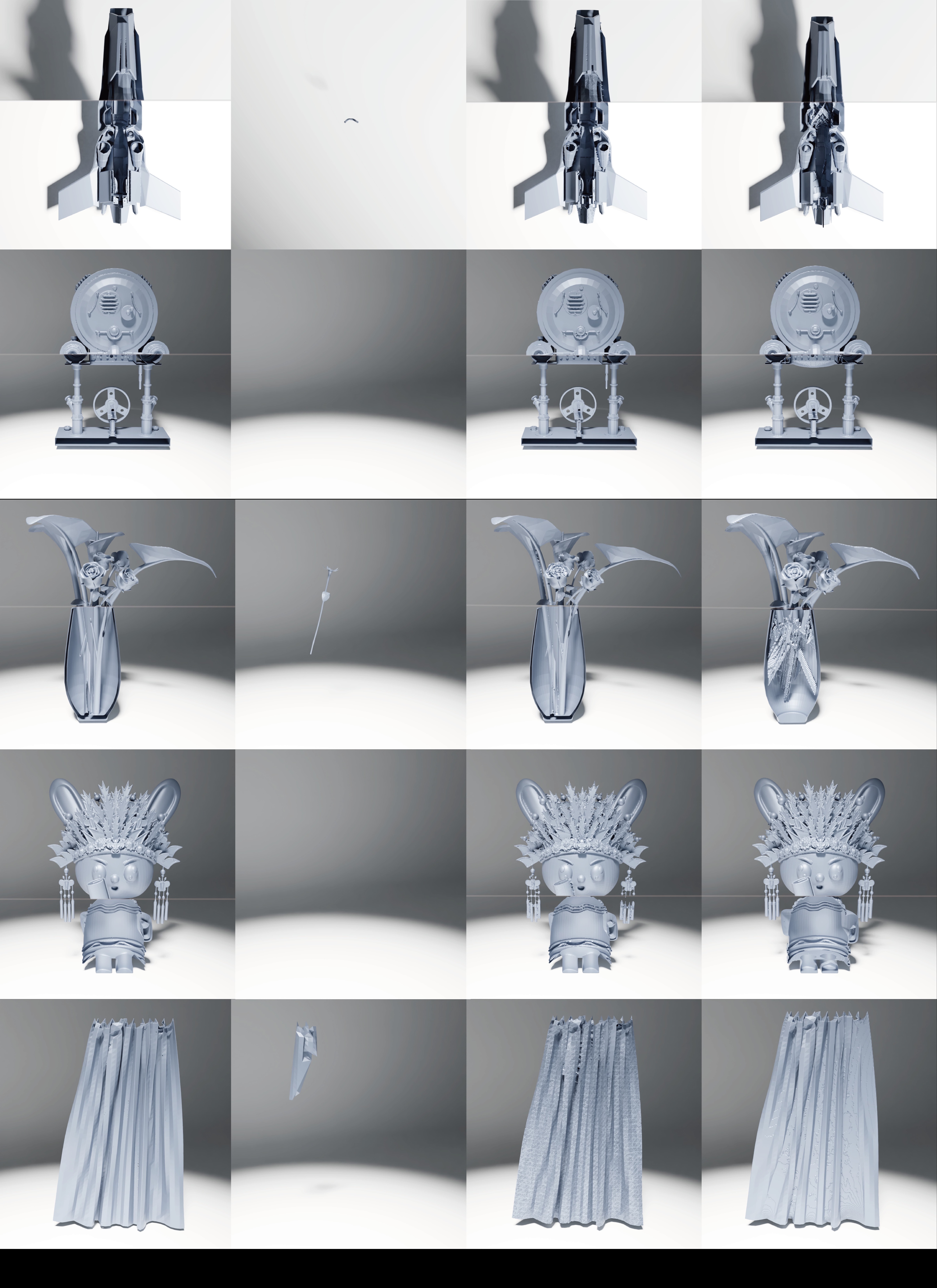}

\caption{\textit{Additional qualitative comparison results of watertight remeshing.}
For each example, we show the reconstructed surface(the upper region) together with a representative longitudinal section(the lower region) if the mesh can be cut.}

  \label{fig: Qualitative results2} 
\end{figure}

\begin{figure}[!htbp]
\centering
  \includegraphics[width=0.9\linewidth,height=0.82\textheight,keepaspectratio]{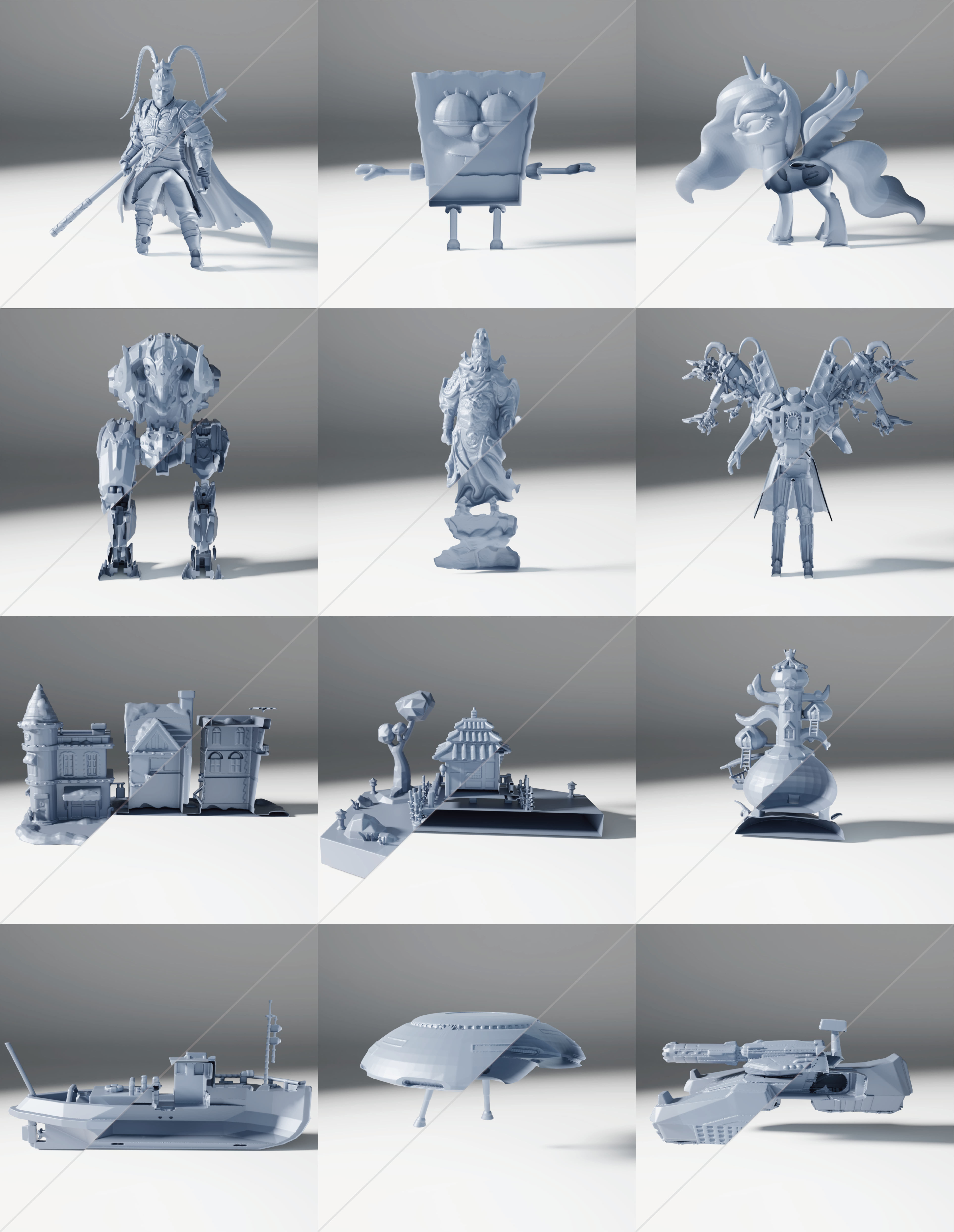}

\caption{\textit{More results of our method.}
For each example, we show the reconstructed surface(the upper triangular region) together with a representative longitudinal section(the lower triangular region).}

  \label{fig: Qualitative results3} 
\end{figure}

To complement the quantitative analyses above, we provide additional qualitative comparisons and reconstruction results that are omitted from the main paper due to space limitations.
Figure~\ref{fig: Qualitative results2} presents further comparisons on challenging inputs, showing that CelloCut consistently produces watertight and geometrically faithful reconstructions.
Figure~\ref{fig: Qualitative results3} presents more results of our method on diverse shapes, further illustrating its robustness across a wide range of geometric structures and topological defects.

\end{document}